\documentclass[aps,prb,twocolumn,groupeaddress]{revtex4}

\usepackage{graphicx}
\usepackage{dcolumn}

\begin{document}


\title{
Circuit with small-capacitance high-quality Nb Josephson junctions
}


\author{Michio Watanabe\footnote{Present address: National Institute of 
Standards and Technology, Division~817, 
325 Broadway, Boulder, CO 80305; electronic mail: watanabe@boulder.nist.gov}}
\affiliation{
Frontier Research System, RIKEN, 2-1 Hirosawa, Wako, 
Saitama 351-0198, Japan
}
\author{Yasunobu Nakamura and Jaw-Shen Tsai}
\affiliation{
NEC Fundamental Research Laboratories, 34 Miyukigaoka, Tsukuba, 
Ibaraki 305-8501, Japan\\
and Frontier Research System, RIKEN, 2-1 Hirosawa, Wako, 
Saitama 351-0198, Japan
}


\date{September 19, 2003}

\begin{abstract}
We have developed a fabrication process 
for nanoscale tunnel junctions 
which includes focused-ion-beam etching 
from different directions.   
By applying the process to a Nb/(Al--)Al$_2$O$_3$/Nb trilayer, 
we have fabricated a Nb single-electron transistor (SET),   
and characterized the SET at low temperatures, $T=0.04-40$~K.  
The superconducting gap energy and the transition 
temperature of the Nb SET agree with the bulk values, 
which suggests high quality Nb junctions.  
The single-electron charging energy of the SET is 
estimated to be larger than 1~K.\\   
\\
Appl.\ Phys.\ Lett. {\bfseries 84}, 410 (2004) [DOI: 10.1063/1.1640798]
\end{abstract}


\maketitle

Circuits with Josephson tunnel junctions 
are one of the most promising candidates 
for quantum bits (qubits) for quantum computation 
in solid-state electronic devices.\cite{Mak01,Fit02}  
The Josephson junctions (JJs) for qubits should 
have large subgap resistance and an appropriate 
$E_J/E_C$ ratio, where $E_J$ is the Josephson energy 
and $E_C$ is the charging energy of the junction.  
The desired range of $E_J/E_C$ depends on  
the degree of freedom (charge, flux, or phase) to be controlled 
in the circuit.   
For cases of charge qubit\cite{Nak99} 
and flux qubit,\cite{Chi03} a typical junction 
size is on the order of $0.1\times0.1$~$\mu$m$^2$, for which 
the fabrication technique of Josephson tunnel junctions 
has been well established only for Al/Al$_2$O$_3$/Al junctions.  
However, materials with larger superconducting gap energy $\Delta$, 
e.g., Nb ($\Delta_{\rm Nb}/\Delta_{\rm Al}\approx8$), 
are more attractive especially for flux qubits, where one 
wants to have large $E_J/E_C$ and keep $E_C$ sufficiently 
larger than the thermal energy $k_BT$.  
Note that $E_J$ is proportional to $\Delta$.  

There have been a number of attempts to 
fabricate small-capacitance Al/Al$_2$O$_3$/Nb 
or Nb/(Al--)Al$_2$O$_3$/Nb JJs.\cite{Mar90,Har94,Pav99,Kim02,Dol02,Dol03}  
Conventional nanoscale fabrication processes based on 
$e$-beam lithography and multiangle shadow evaporation, 
which work fine for Al/Al$_2$O$_3$/Al JJs, 
tend to deteriorate the quality of Nb, 
i.e., $\Delta$ and the superconducting temperature $T_c$ 
are reduced considerably (a summary of this can be found in Fig.~1 
of Ref.~\onlinecite{Dol03}).  
Rather large values of $\Delta_{\rm Nb}/e$ have been obtained 
in single junctions fabricated by a sloped-edge technique\cite{Mar90} 
($1-1.2$~mV) and in a single-electron transistor (SET) 
fabricated by a multilayer technique\cite{Pav99} (1.35~mV).  
However, it would be difficult to fabricate multijunction devices 
using the sloped-edge technique, and the charging energy 
of the SET fabricated by the multilayer technique was 
$\approx 0.15$~K, 
which might be acceptable for flux qubits but would be 
too small for charge qubits and for 
most single-electron-tunneling devices.    

\begin{figure}
\includegraphics[width=0.9\columnwidth,clip]{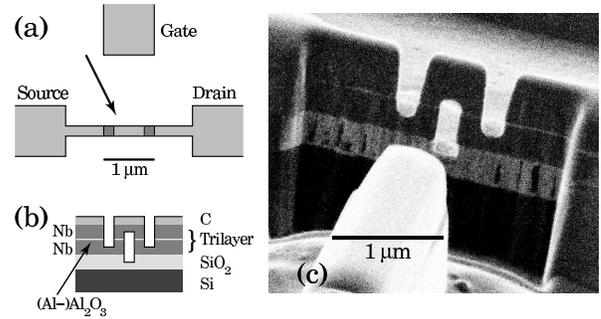}
\caption{\label{fig:sample}
Nb single-electron transistor.  
(a) Top view.  
(b) Side view.  
(c) Focused-ion-beam secondary-electron image 
taken from the direction shown in (a).  
}
\end{figure}

We have developed a process to fabricate high-quality 
small-capacitance Nb/(Al--)Al$_2$O$_3$/Nb JJs 
which includes focused-ion-beam (FIB) etching 
from two different directions.  By employing the process, 
we have fabricated SETs with a three-dimensional structure, 
shown in Fig.~\ref{fig:sample}.  
Details of the process are as follows.  A trilayer
of Nb (thickness: 0.3~$\mu$m), 
Al (0.01~$\mu$m)--Al$_2$O$_3$ ($\approx$1~nm), 
and Nb (0.3~$\mu$m) was sputtered onto a SiO$_2$/Si substrate 
in a single vacuum cycle.  
The Al$_2$O$_3$ layer was formed 
by oxidizing the surface of Al. 
The film was then patterned with standard photolithography 
and Ar$^+$ milling, where the smallest feature size 
in this step was $\approx$5~$\mu$m.  
The dimensions of the trilayer were decreased 
further in a Ga$^+$ FIB system, which has   
three functions: deposition, etching, and observation.  
From the direction perpendicular to the substrate, 
we first deposited C (thickness: $\approx$0.1~$\mu$m) 
by decomposing phenanthrene (C$_{14}$H$_{10}$) gas 
with a Ga$^+$ beam current of 48~pA.  
The role of the C layer is to minimize damage 
of the trilayer by Ga$^+$ during the etching process described below.  
Then, we etched into the pattern shown 
in Fig.~\ref{fig:sample}(a) except for the fine structure 
on the narrow track between the source and the drain.  
For etching, we used two beam currents,  
1.3~nA (rough etching) and 9~pA, not only for 
efficiency but also to minimize the amount 
of material redeposited on the sides of 
the narrow track.    
After tilting the substrate $\approx$88$^{\circ}$, 
three holes were made with a 9~pA beam current on the side of 
the track as shown in Fig.~\ref{fig:sample}(b).  
A double-junction structure formed between the holes.  
An example of the SET fabricated in the above process 
is shown in Fig.~\ref{fig:sample}(c), 
which is a secondary-electron image 
taken in the same FIB system.  
The final step in our fabrication process is 
anodization, and there are two reasons for this.  
One is to eliminate completely short circuit 
of the junction 
due to conducting materials (mostly Nb) 
redeposited during FIB etching, by changing 
the conducting materials into insulating oxides. 
The other is to suppress the contribution 
of the sample surface, 
which might have been damaged by Ga$^+$, 
to electrical conduction.  
Because anodization also reduces the effective area 
of the junctions, we designed the initial junction size 
delineated by the FIB to be $>0.2\times0.2$~$\mu$m$^2$, 
although we could fabricate much smaller junctions 
by this technique.  
We kept the anodization current constant at 
a value in the range 1--10~$\mu$A/cm$^2$ and, for the sample 
discussed below, we anodized up to 60~V, which reduced 
the effective junction area to $<0.1\times0.1$~$\mu$m$^2$.  
The FIB etching was quite reproducible, although 
the resistance of the SET after anodization 
varied from sample to sample.  

\begin{figure}
\includegraphics[width=0.95\columnwidth,clip]{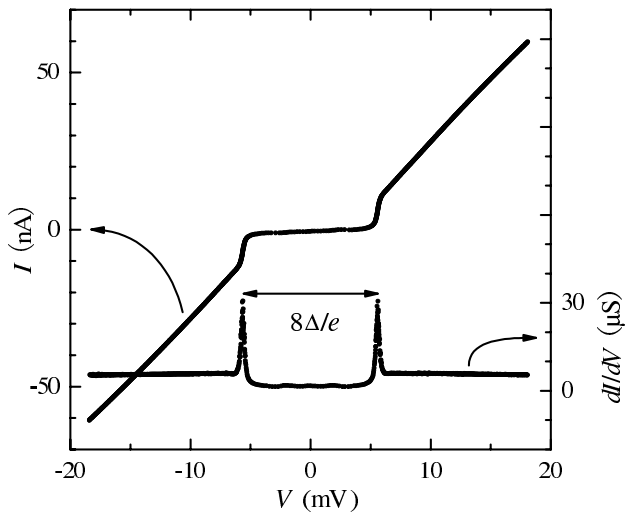}
\caption{\label{fig:IV}
Current-voltage characteristics (upper data set) 
and the differential resistance vs bias voltage $V$ 
(lower data set) 
of the Nb single-electron transistor at $T=3.2$~K.  
}
\end{figure}

\begin{figure}
\includegraphics[width=0.95\columnwidth,clip]{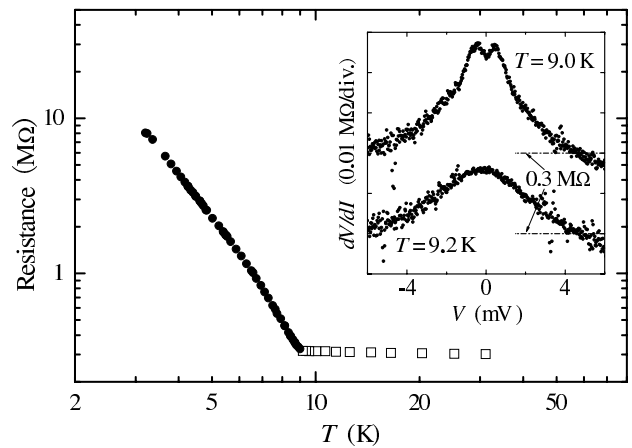}
\caption{\label{fig:RvsT}
Temperature dependence of the zero-bias resistance 
at $T\geq9.2$~K (open squares) and that of the subgap 
resistance at $T\leq9.0$~K (closed circles).   
Inset: Differential resistance vs bias voltage at 
$T=9.0$ and 9.2~K.  The origin of the vertical axis 
is offset for each curve for clarity.   
}
\end{figure}

We measured the samples  
in a $^4$He continuous-flow cryostat 
with a Si-diode thermometer mounted next to the sample holder.  
The current-voltage ($I$-$V$) characteristics of 
one of the samples 
at $T=3.2$~K is shown in Fig.~\ref{fig:IV}.   
The $I$-$V$ curve exhibits a sharp superconducting gap, 
whose value corresponds to two Nb JJs.  
We obtained $\Delta_{\rm Nb}/e=1.4$~mV from 
the distance between the peaks in the differential conductance 
$dI/dV$ vs $V$ curve (see the lower data set of Fig.~\ref{fig:IV}).  
Here the charging energy of the SET is negligibly
small for the estimate of $\Delta_{\rm Nb}$ 
as will be discussed later.  
This value of 1.4~mV is the largest   
for $\sim0.1\times0.1$~$\mu$m$^2$ junctions.    
Moreover, it agrees with the values at liquid $^4$He 
temperatures for 1--10~$\mu$m scale JJs fabricated by 
a standard photolithographic technology for integrated JJ circuits.  
We also measured the temperature dependence 
of the $I$-$V$ curve in order to determine $T_c$.  
The $I$-$V$ curve looks almost linear at $T>8$~K.  
However, when we look at the differential resistance $dV/dI$, 
it is easy to find a qualitative difference 
between the $dV/dI$ vs $V$ curves at $T\leq9.0$~K and those at 
$T\geq9.2$~K.  The inset of Fig.~\ref{fig:RvsT} shows    
the $dV/dI$ vs $V$ curve at $T=9.0$ and 9.2~K.  
The curve at 9.0~K has a dip in the middle, 
while that at 9.2~K does not.  
Based on this qualitative difference, we determine that 
$T_c=9.1\pm0.2$~K, which agrees with the bulk value of 9.2~K 
within the error.  Here $\pm$0.2~K includes 
all errors in thermometry and $T_c$ determination.    
We confirm the adequacy of the above $T_c$ determination  
in Fig.~\ref{fig:RvsT} by plotting versus $T$ 
the zero-bias resistance for $T\geq9.2$~K (normal state) 
and the subgap resistance for $T\leq9.0$~K 
(superconducting state), where we have defined 
the subgap resistance as the maximum value of $V/I(V)$.  
From the values of $\Delta$ and $T_c$, we conclude that 
the quality of our JJs is high and it has not deteriorated 
in the FIB etching process.  

Let us look at the properties of the sample 
as an SET.  
The gate electrode is located $>$1~$\mu$m from 
the double-junction system (Fig.~\ref{fig:sample}), 
and the voltage $V_g$ applied to the gate 
modulates the current even up to 5~K (data not shown).  
Thus, the sample indeed works as an SET.  
In order to characterize 
the sample further, we cooled it down to 0.04~K in a 
$^3$He--$^4$He dilution refrigerator.  
At $T\leq1$~K, the subgap resistance becomes 
$>$10$^2$~M$\Omega$, which is significantly larger 
than the normal-state resistance, 0.3~M$\Omega$, 
and again, suggests high quality junctions.     

\begin{figure}
\includegraphics[width=0.95\columnwidth,clip]{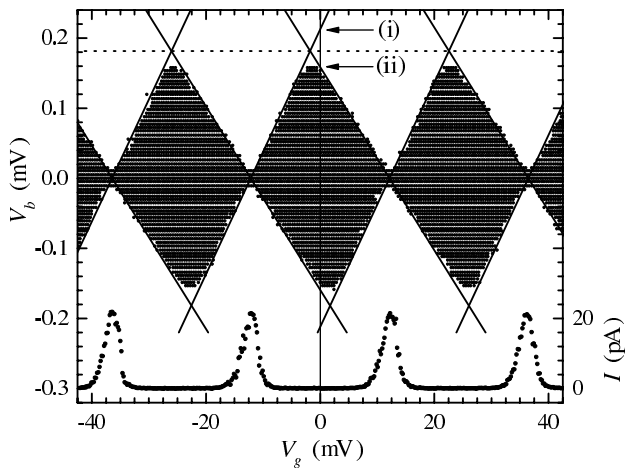}
\caption{\label{fig:diamond}
Blockade of single-electron tunneling 
in the normal state ($B=2.7$~T) at $T=0.05$~K.  
Lower data set: Current $I$ vs gate voltage $V_g$ 
at $V=0.026$~mV, where $V$ is the bias voltage.  
Upper data set: Region of $|I|<10$~pA on the $V_g$-$V$ 
plane.  The closed diamonds indicate the 
Coulomb-blockade region.  
The horizontal dotted line 
corresponds to $2E_C/e$, where $E_C$ is the 
single-electron charging energy of the SET.       
}
\end{figure}

The parameters of a SET, such as the capacitances 
$C_1$ and $C_2$ of the tunnel junctions and the capacitance 
$C_g$ between the island electrode and the gate electrode, 
can be determined by measuring the blockade of single-electron 
tunneling in the normal state 
at a low enough temperature.\cite{Ing92}    
We drove the Nb SET into the normal state by applying 
a magnetic field of 2.7~T perpendicular to the substrate, 
and measured $I$-$V_g$ curves 
for different values of $V$ at $T=0.05$~K.   
An example of the $I$-$V_g$ curve, 
which is for $V=0.026$~mV, is shown 
in the lower data set of Fig.~\ref{fig:diamond}.  
One period in the $V_g$ axis corresponds to $e/C_g$, 
so that $C_g=7$~aF, which is consistent with the geometry.  
In the upper data set of Fig.~\ref{fig:diamond}, 
we estimate the zero-current region at $T=0$ 
on the $V$-$V_g$ plane (Coulomb diamond).  
The horizontal dotted line 
corresponds to $2E_C/e$ of the SET.  
Thus, from $2E_C/e=0.18$~mV in Fig.~\ref{fig:diamond}, 
we obtain $E_C/k_B=1.1$~K, which is 
consistent with the effective junction area 
of $<0.1\times0.1$~$\mu$m$^2$ 
and sufficiently larger 
than the base temperature of a typical 
dilution refrigerator, $<$0.1~K.  
When $C_g\ll C_1,$ $C_2,$ 
which is the case in our Nb SET, 
$C_1$ and $C_2$ are estimated from the slopes 
of the solid lines in Fig.~\ref{fig:diamond}.  
We find that $C_1/C_2=0.75$, although we designed 
it so that $C_1=C_2$.  

The reason for asymmetry in the junction capacitance 
is that we had to anodize the SET up to a large voltage 
in order to eliminate the short circuit.  
The accuracy of the capacitances could be 
improved by introducing a step of reactive-gas-assisted 
etching at the end of the FIB process.  
We have already confirmed that XeF$_2$ gas enhances 
the etching rate of Nb $\approx$10$^2$ times 
and ``cleans up" redeposited Nb.  
By introducing the step, we would be able to reduce 
the anodization voltage considerably.  
Reducing the anodizaiton voltage would also improve 
the yield of the fabrication process.    

It should be noted that our process is also applicable 
to more complex circuits.  Moreover, it is much more 
flexible in terms of circuit pattern 
than the conventional technique based on $e$-beam 
lithography and shadow evaporation, or the multilayer 
technique in Ref.~\onlinecite{Pav99}.  
Very recently, a similar FIB-etching technique was 
independently developed, and nanoscale single junctions 
with a variety of materials were fabricated.~\cite{Bel03}  
Thus, the process is not limited to 
Nb/(Al--)Al$_2$O$_3$/Nb junctions.  

In the superconducting state, one expects that 
the supercurrent flowing through the SET 
depends on $V_g$ periodically\cite{Joy95} and 
that the period is $2e/C_g$ ($2e$ periodic).   
In many experiments, however, a period of 
$e/C_g$ ($e$ periodic), 
which suggests the existence of subgap quasiparticle 
states,\cite{Joy95} has also been seen.  
In small-capacitance Al/Al$_2$O$_3$/Nb 
or Nb/(Al--)Al$_2$O$_3$/Nb systems, 
only $e$ periodicity has been reported.  
In our Nb SET, the measured supercurrent at $T=0.04$~K and $B=0$ 
is also $e$ periodic, and its magnitude is on the order of 10~pA, 
which is $\approx$10$^{-3}$ the theoretical maximum, 
$\sim I_0/2$, where $I_0\equiv \pi\Delta/2eR_n$ is the 
Ambegaokar-Baratoff critical current and $R_n$ is 
the normal-state resistance of the junction.  
Further investigation of the periodicity by measuring 
high-quality Nb SETs with different parameters would 
clarify whether the $e$ periodicity is intrinsic to Nb or not.  

In summary, we have fabricated a high-quality 
Nb SET with $E_C/k_B>1$~K by developing a fabrication process 
for nanoscale tunnel junctions.  
The process is much more flexible 
than conventional ones based on $e$-beam lithography.   

The authors are grateful to Y. Kitagawa for preparation of 
the Nb/(Al--)Al$_2$O$_3$/Nb trilayer, to M. Ishida 
for assistance with the FIB system, 
to H. Akaike for valuable comments on anodization, 
and to Yu. Pashkin, T. Yamamoto, and O. Astafiev 
for fruitful discussions. 
This work was supported in part 
by the Special Postdoctoral Researchers Program 
of RIKEN and by MEXT.KAKENHI (Grant No.~15740190).  


\end{document}